\documentclass[ fleqn,10pt]{wlscirep}

\usepackage[utf8]{inputenc}
\usepackage[T1]{fontenc}
\usepackage{dsfont}
\usepackage{bm}
\usepackage{hyperref}
\usepackage{tikz}
\usepackage{xcolor}
\title{Truly local topological dynamics of driven defects in Chern insulator}

\author[1, 2, *]{D. B. Golovanova}
\author[1, 2]{A. R. Yavorsky}
\author[1]{A. A. Markov}
\author[1, 3]{A. N. Rubtsov}
\affil[1]{Russian Quantum Center, Moscow 121205, Russia}
\affil[2]{Moscow Institute of Physics and Technology, Dolgoprudny, 141701, Russia}
\affil[3]{Lomonosov Moscow State University, Moscow 119991, Russia}

\affil[*]{golovanova.db@phystech.edu}

\begin{abstract}
 Robust zero modes supported by defects is one of the key features of topological matter. Its presence renders a system topologically inhomegeneuous, thus having no well-defined global topological invariant. The quantities labeling different areas of the sample according to their topological state were dubbed local topological markers. Here we study their dynamics and the possibility to control their distribution over the sample. We suggest a new perspective on the evolution of local markers. It gives a clear physical description of the markers evolution in terms of response functions and the ease of measurement. Furthermore, new markers' equations of motion are truly local, being ensured that the current of the marker exists and obeys the lattice continuity equation. The formalism presented does not rely on the single-particle quantities therefore might be extended to interacting systems.
\end{abstract}
\begin{document}

\flushbottom
\maketitle
%
%
\thispagestyle{empty}

\section*{Introduction}

Formation of robust gapless modes in the presence of defects is a defining property of the topological matter \cite{halperin1982quantized,teo2010topological}. Moreover, this property shows a great promise with regards to applications of topological materials, such as dissipationless lines \cite{buttiker1988absence}, spintronics, and quantum computation\cite{qi2011topological}. Therefore, the ability to control the presence and properties of defects is of both fundamental and practical interest. 

 \begin{figure}[h!]
 \centering
	\includegraphics[width=0.8\linewidth]{"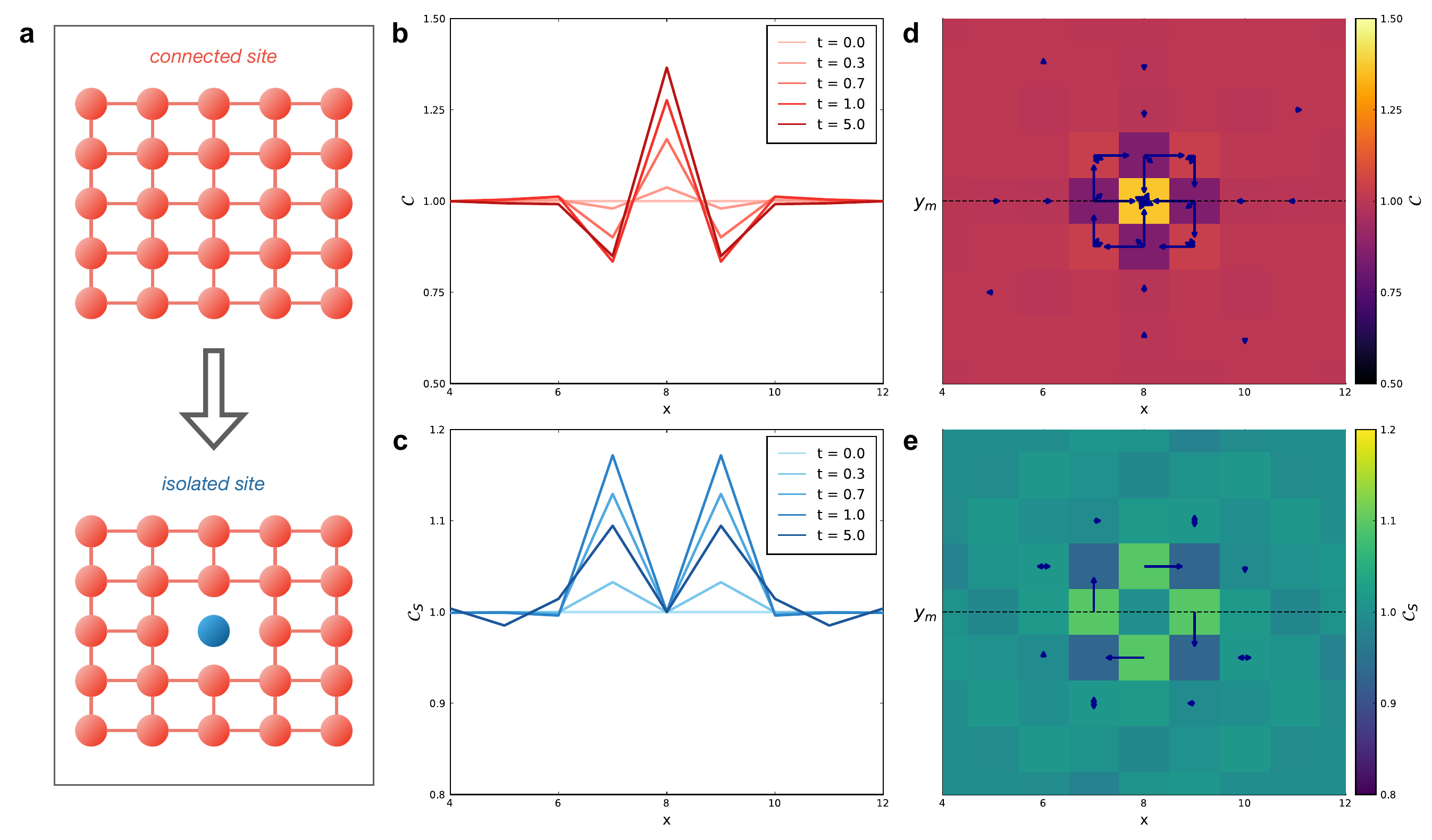"}
	\caption{\textbf{The dynamics of the LCM under quench isolating a single site in the sample’s middle.} \textbf{a} The scheme of the "cutt-off" scenario. At the t=0 the hoppings from the central site are set to zero. Simultaneously the on-site parameters are changed.  \textbf{b} Teleportation of the marker value from the central site which was cut off. The evolution of the LCM distribution is described by Eq. ~\ref{eq:chern_m} as in Refs.\cite{d2015dynamical,privitera2016quantum, caio2019topological, toniolo2018time} . The insert shows the behaviour of $C(\mathbf{r},t)$ non-local currents close to the site. Shown here is one of the many possible non-local currents described by Eq. ~\ref{eq:chern_m}, for more details see Supplementary materials. \textbf{c} Distribution of $C(\mathbf{r},t)$ Eq. ~\ref{eq:nq_streda} at different times. The blue arrows on insert shows the absence of LCM currents from the isolated site.}
	\label{fig:comp_lcm}
\end{figure}
Samples with defects are inhomogeneous and require special care from a theoretical perspective. Global topological indices, e.g. Chern number \cite{thouless1982quantized}, are not applicable directly to the systems with topological defects, as they characterise the whole system. Recently, a family of quasitopological quantities, coined local topological markers, has been developed and studied \cite{kitaev2006anyons,bianco2011mapping, loring2011disordered}. They provide an estimation of the global topological index based on the parameters of the neighborhood of a given point in space and are not necessarily strictly quantized. The robust characteristic is the average of the markers over large areas of a system \cite{bianco2014chern}. 


 
 Out of equilibrium systems reveal another striking difference between the global and local topological quantities. Global indices of infinite systems are conserved during unitary evolution unless the symmetry of the problem changes dynamically \cite{toniolo2018time,caio2015quantum,mcginley2019classification}. In finite samples, characterised by local markers, the situation is different. The topological index - an average of local markers over large areas - may change as was demonstrated by numerical simulations \cite{d2015dynamical,privitera2016quantum, caio2019topological} and analytically \cite{toniolo2018time}.
 
Neither local topological markers\cite{kitaev2006anyons,bianco2011mapping, loring2011disordered}, nor their equations of motions are unique. It would be highly desirable that the dynamics for the markers would be local. That is the marker would flow from a site to its neighborhood, not teleport to distant sites. In an insightful paper \cite{caio2019topological} it was conjectured that the dynamics of the marker satisfy this property and governed by local currents emanating from the system's boundaries, defined implicitly through the lattice continuity equation on the marker. We demonstrate that such currents could not be defined without the modification of equations of motions for the markers used before \cite{d2015dynamical,privitera2016quantum, caio2019topological, toniolo2018time} . Most clearly it can be seen in the numerical experiment shown in Fig.\ref{fig:comp_lcm} \textbf{b}. There the teleportation of the marker from the disconnected central site is shown. 

In the present manuscript we suggest a new physical perspective on the dynamics of the local topological markers. We have used the connection of markers to the charge response to a magnetic field. New equations of motion have a clear physical meaning, do obey the continuity equation, and are much easier to measure in an experiment as require only density measurements. Furthermore, our approach does not rely on single-particle projectors onto filled states. Therefore it might be used in the interacting systems.

Equipped with the new approach, we numerically study the possibility to control the position of line defects in a finite 2-d Chern insulator. For a start, we studied a predictable setting depicted schematically in Fig.\ref{fig:setting}. The system is prepared so that the central region is in a topologically nontrivial phase and surrounded by the trivial insulator, as indicated by the initial distribution of the local topological marker.  We considered the dynamics of local markers and their currents, corresponding to the slow movement to the right of the topological domain. We demonstrate the possibility of quasiadiabatic control of topological marker distribution in the system. We show that the steady-state distribution of the local marker catches the location of gapless modes. 

\section*{Local Marker Dynamics}



\textit{Local Chern marker.} Two dimensional non-interacting systems are classified topologically by Chern number \cite{thouless1982quantized} $C$ in the absence of symmetries \cite{altland1997nonstandard} to protect. Strictly speaking, topological phases with $C\neq 0$ realize in the thermodynamic limit on a torus. Real-world samples subjected to open boundary conditions necessarily contain topological defects; one can be certain that at least the boundary vacuum - Chern insulator is present in the system. Further, the sample may contain macroscopically large patches with parameters corresponding to different Chern numbers. One needs a substantially local quantity that allows to label different parts of the systems according to their Chern number to work in such settings. Local topological markers \cite{kitaev2006anyons,bianco2011mapping, loring2011disordered} do precisely the job.

Physically the markers may be connected to the charge response to applied magnetic \cite{bianco2013orbital,loring2011disordered,mitchell2018amorphous, kitaev2006anyons} or electric \cite{peru} field.  For concreteness, we will use the Local Chern Marker\cite{bianco2011mapping} (LCM). \footnote{The Bott index's\cite{loring2011disordered}  application to the systems with zero-modes is problematic. Much of the discussion to follow can be applied as well to the Kitaev's marker closely related to the LCM both formally \cite{bianco2011mapping} and physically \cite{mitchell2018amorphous}}. For this marker the local version of the Streda formula \cite{streda1982theory} holds\cite{bianco2013orbital} up to insignificant corrections appearing on the system borders or defects: 

 \begin{equation}  \label{eq:chern_streda}
 \mathrm{C_S}(\mathbf{r})= \phi_0\frac{\delta n(\mathbf{r})}{\delta B_z},
\end{equation} 

here $\phi_0=$ is the flux quanta. The variation of the average density $n(\mathbf{r})$ is taken with respect to a uniform magnetic field which is supposed to be turned on adiabatically slow. The Kitaev marker \cite{kitaev2006anyons} can be obtained from a very similar equation with the uniform magnetic field replaced by a local one \cite{mitchell2018amorphous}. 

 A generic example of the marker distribution in an inhomogeneous sample is presented in Fig.~\ref{fig:setting}\textbf{b}. The marker is very close to quantized value in the ''bulk'' of a patch in a trivial or topological phase. In vicinity of defects LCM develops sharp negative peaks. A physical intuition for this behaviour is that defects serve as electron sources for the currents flowing towards the bulk when a small magnetic field is applied.

The definition of the markers is typically \cite{bianco2013orbital,loring2011disordered, kitaev2006anyons} given in terms of the  projectors $\hat{P}$ onto the single-particle filled states. The LCM in particular is defined as the spatially resolved commutator of  $\hat{X}$ and $\hat{Y}$ operators dressed in projectors:

\begin{equation}  \label{eq:chern_m}
 C(\mathbf{r})=  - 2\pi i  \sum_{\alpha}\left\langle\mathbf{r_\alpha}|\left[\hat{P} \hat{X} \hat{P}, \hat{P} \hat{Y} \hat{P}\right]| \mathbf{r_\alpha}\right\rangle.
 \end{equation}

This the formula highlights two more of the marker's properties. First, it is a quasi-local quantity due to exponential decay of $\hat{P}(\bm{r},\bm{r}')$ with the distance between $\bm{r}$ and $\bm{r}'$ \cite{lieb1972finite}. Second, it sums over a finite sample to zero, as a commutator is traceless.

\textit{LCM dynamics} 

It is very tempting to use the markers to understand the topological properties out of equilibrium. How the time-dependent markers should be defined is not immediately clear, however. The markers are not strictly speaking operators, obeying the Heisenberg equation of motion. Rather they are operator-valued functions of the system's state. An appealing approach is to use the same function (e.g. Eq. \ref{eq:chern_m} for LCM) of the state and allow it to evolve \cite{privitera2016quantum,toniolo2018time,caio2019topological,d2015dynamical}. It was revealed that in finite-size systems the markers can change \cite{d2015dynamical,privitera2016quantum}, in contrast to global topological indices \cite{d2015dynamical,caio2015quantum,toniolo2018time}. Importantly, their evolution reflects the change in a topological phase \cite{d2015dynamical,privitera2016quantum, caio2019topological}.

A remarkably simple and appealing picture of the dynamics of the LCM was presented in Ref. \cite{caio2019topological}. It was conjectured that the marker's evolution is governed by local currents emanating from the edges of a sample. The currents were defined through a lattice continuity equation. The claim is highly non-trivial, as the LCM is not a fully local quantity. Let us stress that, the global conservation of the LCM - $\sum C(\bm{r})=0$ - does not guarantees the existence of a local continuity equation. In fact, such a continuity equation can not be satisfied by $C(\mathbf{r},t)$ in general. 

Most clearly it can be observed in the following numerical experiment, shown in Fig.~\ref{fig:comp_lcm}\textbf{(a)}. Consider a homogeneous sample in the topological phase. At the time $t=0$ a central site is cut off from the rest of the system. That is all the hoppings from it are quenched to zero. Simultaneously, the on-site parameters are changed. Were $C(\mathbf{r},t)$ satisfy a lattice continuity equation, no change of the marker would be observed on the site.

The Fig. \ref{fig:comp_lcm}(c) demonstrates that the on-site value of the marker does change. The time-dependent Eq.~\ref{eq:chern_m} predicts the teleportation of the marker from the disconnected site. Therefore, the numerical evidences \cite{caio2019topological} of the current-like local behaviour of the LCM are seeking an explanation and elaboration. We demonstrate in the Supplementary how the non-local currents in this case may be explicitly defined.        

We suggest a new perspective on the markers' evolution, generalizing the Eq.~\ref{eq:chern_streda}:

\begin{equation}  \label{eq:nq_streda}
\mathrm{C}_S(\mathbf{r}, t)= \phi_0\frac{\delta n(\mathbf{r}, t)}{\delta B_z},
\end{equation} 
here the magnetic field is supposed to be turned on adiabatically at $t=-\infty$. 

Serving as good topological marker as the defined by means of Eq.~\ref{eq:chern_m}, the new approach has important advantages. First, this formula has a clear physical interpretation as a spatial resolution of the Hall conductivity. Second, it is easier to measure.  The typical protocol of reconstructing the LCM requires the knowledge of off-diagonal elements of the projectors $P(\bm{r},\bm{r}')$ \cite{caio2019topological,irsigler2019microscopic}. This is a much harder experimental task \cite{ardila2018measuring}, then the density measurements \cite{bakr2009quantum}. Third, the evolution of the LCM as described by the Eq.~\ref{eq:nq_streda} satisfies the lattice continuity equation:

\begin{equation} \label{eq:cont_eq}
\begin{split}
 \partial_{t} \mathrm{C}_S(\mathbf{r}, t)=-\text{div} \mathbf{J}^c ,\\
J^c(\mathbf{r},\bm{r}', t) = \frac{\delta J^e(\mathbf{r},\bm{r}', t)}{\delta B_z}.
\end{split}
\end{equation}

where $J^e(\mathbf{r},\bm{r}', t)$ is the average current on the bond between sites $\bm{r}$ and $\bm{r}'$. 

Therefore the Eq.~\ref{eq:nq_streda} forces the dynamics to be local, as we can see both numerically (Fig.~\ref{fig:comp_lcm}\textbf{(c)}) and analytically. Moreover, it provides a physical explanation why the current-like behaviour of the LCM was observed\cite{caio2019topological} in the first place. The reason is that the marker currents can be directly connected to the electric currents as Eq.~\ref{eq:cont_eq} demonstrates. See Supplementary for the further discussion of the quench dynamics as in Ref.~\cite{caio2019topological}.

\section*{Controlling an inhomogeneous topological lattice.}
	\begin{figure}[h!]
		\centering
		\includegraphics[width=0.9\linewidth]{"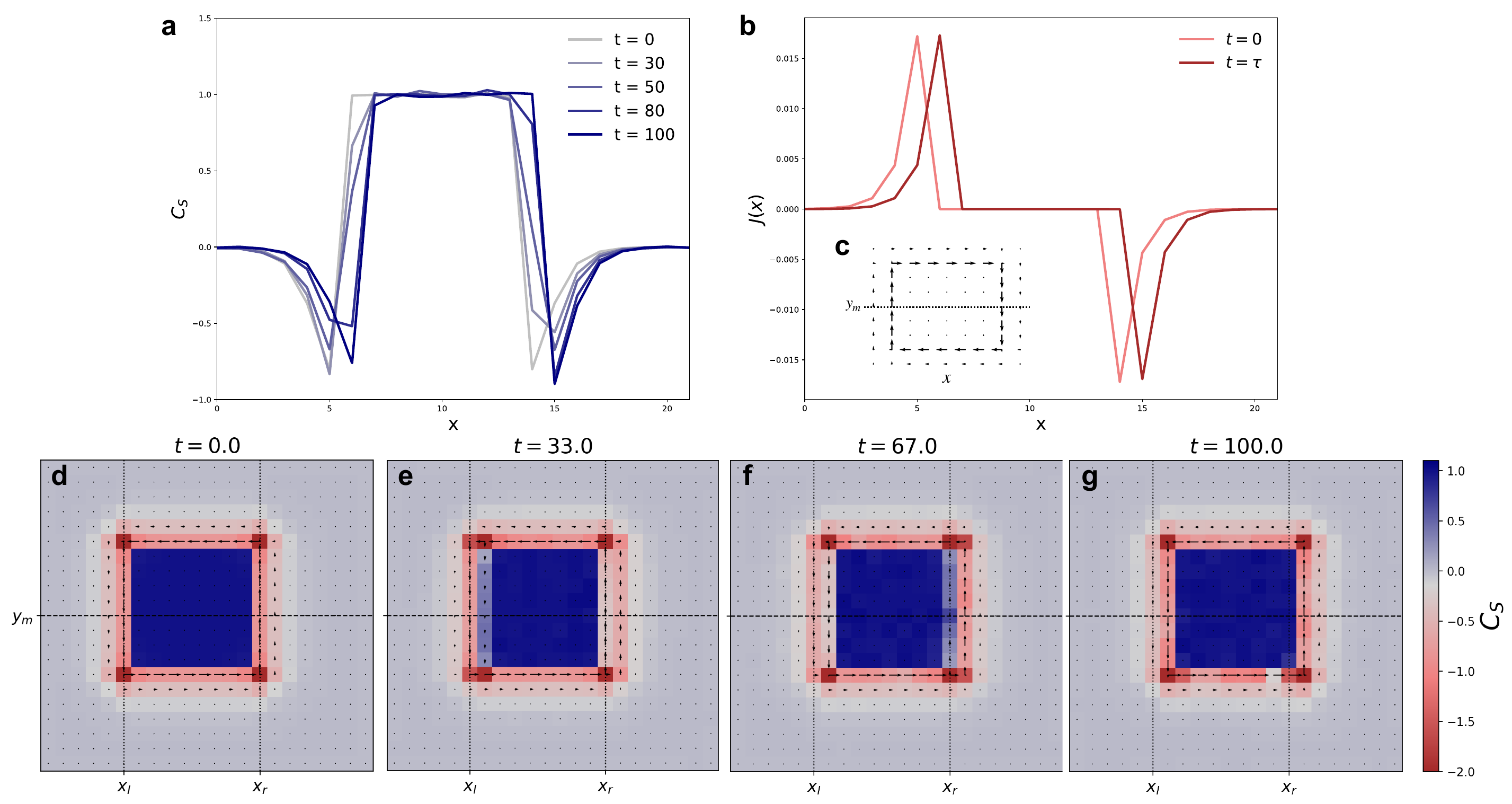"}
		\caption{\textbf{The movement of the “topological” area inside the Chern insulator in the trivial phase.} \textbf{a} The evolution of LCM corresponding to one-site shift of the area in the slice indicated in \textbf{c}. \textbf{b} Corresponding evolution of $y$ component of electric current in units of $et_ha/\hbar$. The current was obtained after shifting of chemical potential by $0.1\Delta$. \textbf{d}-\textbf{g} The spatial distribution of the LCM and its currents on the lattice at different times. $\tau=100$ $\hbar/t_h$.  Lattice size is $22$ by $21$, topological domain size is $10$ by $10$.
		}
		\label{fig:flow}
	\end{figure}


The control over the spatial distribution of topological properties is a very important problem both fundamentally and practically. The tunable position of robust zero modes is particularly fruitful, as these modes are key to the most impressive possible applications of topological matter \cite{buttiker1988absence, qi2006topological}. From the local marker perspective the task might be stated as the possibility to dynamically control the marker distribution. As a starting point on the road we studied the slow dynamics of the topological markers.

We considered the slow movement of a domain in topological phase inside an otherwise trivial sample as presented in Fig.~\ref{fig:setting}. After a time $\tau$, the domain with ''topological''  parameters shifts by one site to the right under a linear time-protocol (See Methods for the details).  Does a shift in the parameter's distribution mean a real shift of the topological domain? To address the question we calculated the electric currents shifting the position of the chemical potential of the system by $\sim 0.1\Delta$. The steady-state amplitude of the electric currents in the y-direction in the middle of the sample and the distribution of currents on the bonds of the lattice is presented in Fig.~\ref{fig:flow} \textbf{b}-\textbf{c}. The shift of edge currents confirms that topological domain moved one site to the right. 


As the system evolves, the distribution of the marker changes. The distribution of the local Chern marker follows the shift of the topological area, provided that the process of moving was sufficiently slow. The protocol period should be much larger than the inverse of bulk gap $\tau \gg 1/\Delta$. The Fig.~\ref{fig:flow} \textbf{a} demonstrates how the distribution evolves during the time $\tau=100$. In this regime, the steady-state distribution resembles that of the equilibrium ground state transferred as a whole one site to the right. 

The local currents of the marker were observed near the borders of the topological domain. These currents are shown by the arrows in Fig.~\ref{fig:flow} \textbf{d}-\textbf{g}.  The defects played the role of the electron reservoirs for the bulk in the presence of the small magnetic field.


	\begin{figure}
		\centering
		\includegraphics[width=1\linewidth]{"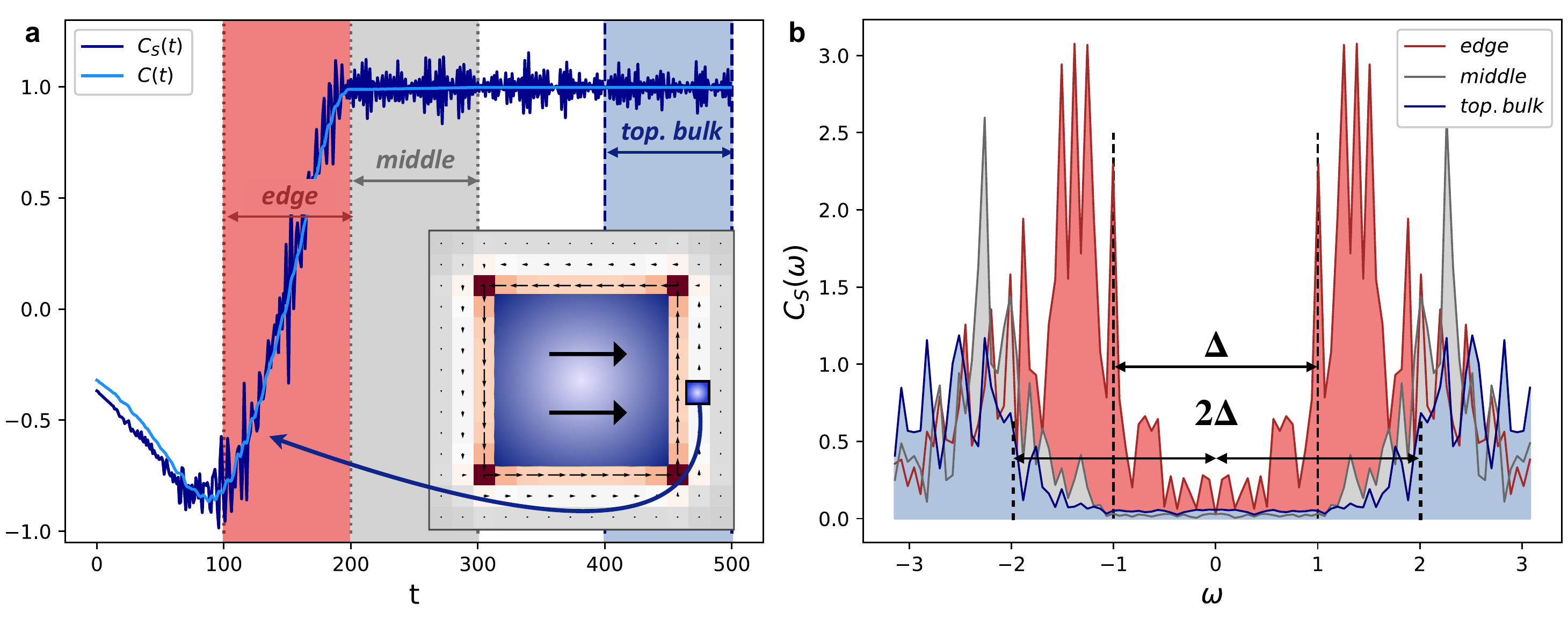"}
		\caption{\textbf{Topological marker behavior on the site in three different modes.} \textbf{a} Time-dependence of LCM on the site $x = x_r+1, \: y=y_m$ highlighted in blue in the insert. The dark blue curve corresponds to $C_S(r,t)$ from the Streda-based formula~\ref{eq:nq_streda}, the light blue $C(r,t)$ to the formula~\ref{eq:chern_m}. The topological domain here moves three sites to the right, during $\Delta t=300$. In the beginning, the site is to the right of the topological domain boundary. At $t=100$ it appears on the boundary, then $t=200$ in an intermediate mode, finally, at $t=300$ in the topological bulk. In the insert the initial position if highlighted. \textbf{b} Fouirer components of $C_S(r,t)$ oscillations in three modes defined on the previous figure. $\Delta$ is the energy gap. Lattice size is $22$ by $21$, topological domain size $10$ by $10$.}
		\label{fig:ch_afr}
	\end{figure}
	
\textit{Oscillations of LCM.} 

We compared the time-dependency of $C_S(t)$ and $C(t)$ (Eq.~\ref{eq:chern_m},\ref{eq:nq_streda}). They agree qualitatively, though differ in details. As it can be seen from Fig.\ref{fig:ch_afr} \textbf{a} the amplitude of oscillations of the first is several times higher than the amplitude of the second. The nature of these oscillations can be explained by the fact that inhomogeneous samples always contain zero modes. The small energy of hamiltonian perturbation excites these levels and makes densities on the border oscillate. The rest of the system also responds to these fluctuations. 

We considered the process of moving a topological domain three sites to the right. In Fig.\ref{fig:ch_afr} \textbf{a} the time-dependence of LCM on the site with coordinates $x = x_r+1, \: y=y_m$ is presented. When the site is on the topological edge at $t=100$, the LCM have maximum in both the value and the amplitude of oscillations. From $100$ to $200$ there is a drift of the marker towards its final value $C=1$ with a constant speed determined by $1/\tau$. The site continues to move into topological bulk from $200$ to $300$ and undergoes free-evolution without the change in Hamiltonian for the time remaining.

The site position determines the frequency of the oscillations and their magnitude. Amplitude-frequency response is presented in the Fig.\ref{fig:ch_afr} \textbf{b} for oscillations in different regimes. In the case when the site is moving from the edge at the time from $100$ to $200$, the dominant frequency coincides with the transition from the zero-energy state to the bottom of the band( Van Hove singularities In the thermodinamic limit). When the site moves into the bulk, only transitions between two energy bands survive. This is shown in the Fig.\ref{fig:ch_afr} \textbf{b}. The oscillations are less pronounced, the further from the border between different phases the site is - another indication of the locality of the dynamics of the marker.

\section*{Discussion}

 We provided a new clear physical picture of the local marker's evolution. We demonstrated that the evolution of the markers can be enforced to be fully local, despite the markers complex quasi-local definition. Local marker currents were directly connected to the local electrical currents. With the new tool at hand, we studied the possibility to control the line defects in a non-homogeneous finite sample. Our work paves the way to the development of tunable devices harnessing topologically protected modes on line defects.


The setting we studied may be realized in an experiment. Most notably, modern cold atomic platforms have the technologies to create and control topological interfaces \cite{goldman2016creating} in a single sample. In our approach, the dynamics of the marker can be traced by density measurements only. Such a technique is available at the moment \cite{bakr2009quantum}. The usual approach requires the knowledge of the single-particle density matrices \cite{irsigler2019microscopic} which is much harder in practice \cite{ardila2018measuring}.


Let us suggest possible extensions of our work. Interacting topological insulators and in particular fractional Chern insulators are very interesting to address with our method. Our equations of motion are ready can be applied to many-body systems as they do not rely on single-particle projectors as usual. In an interacting system the projectors to filled states are not defined, complicating the generalization of the local markers. On the other hand, recent equilibrium calculations indicate that the Streda-based formula Eq.\ref{eq:chern_streda} can be used as a local marker for fractional phases\cite{wang2021measurable}. Another important task is to find the optimal method to control the local marker's distribution. This requires further analytical and numerical studies of the nonhomogeneous topological systems out of equilibrium.

\section*{Methods}
\label{sec:methods}
 
\textit{Model.} We consider a model of Chern insulator ~\cite{qi2006topological} with Hamiltonian:

\begin{equation} \label{eq:Ch_eq}
H(t)=\sum_{n} t_h (c_{n,\sigma}^{\dagger} \frac{\hat{\sigma}_{z}-i \hat{\sigma}_{x}}{2}  c_{n+\hat{x},\sigma} + c_{n,\sigma}^{\dagger} \frac{\hat{\sigma}_{z}-i \hat{\sigma}_{y}}{2}  c_{n+\hat{y},,\sigma} + h. c.) + m(\mathbf{r}, t) \sum_{n} c_{n}^{\dagger} \hat{\sigma}_{z}  c_{n},
\end{equation}
for fermi-particles on a finite $N_x\times N_y$ square lattice with the lattice constant $a=1$. Here, Pauli matrices $\sigma_i$ act in the spin subspace. In the homogeneous case $m(\mathbf{r},t) = const$, the model's phase is determined by the ratio of parameters $m$ and $t_h$:

\begin{equation} \label{m_det}
\begin{array}{l}
-2t_h<m<0 \quad \text{topological;} \quad C=1 \\
0<m<2t_h \quad \text{topological;} \quad C=-1 \\
m<-2t_h, \; m>2t_h \quad \text{trivial;} \quad C=0\\
\end{array}
\end{equation}

The phases are separated by the Dirac semi-metal phases. The energy and time scale are fixed by setting $t_h$ and $\hbar$ to $1$.

 To calculate the LCM according to the Eq.\ref{eq:nq_streda} the density of electrons calculated numerically in the presence The magnetic field is coupled to the system by means of the Peierl's substitution, which introduces a site-dependent phase factor in the hopping matrix $t(\bm{r}_1,\bm{r}_2)$:
	
	\begin{equation}
	t(\bm{r}_1,\bm{r}_2) = t_h\cdot e^{i\frac{2\pi}{\phi_0}\int_{\bm{r}_1}^{\bm{r}_2}\bm{A}(\bm{r})\cdot \bm{dr}},
	\label{eq:hop}
	\end{equation}
	
The variation in Eq.~\ref{eq:nq_streda} is taken with respect to a uniform magnetic field with a flux small compared to the flux quantum. 

\textit{Time protocol.}

 \begin{figure}[h!]
 \centering
	\includegraphics[width=0.8\linewidth]{"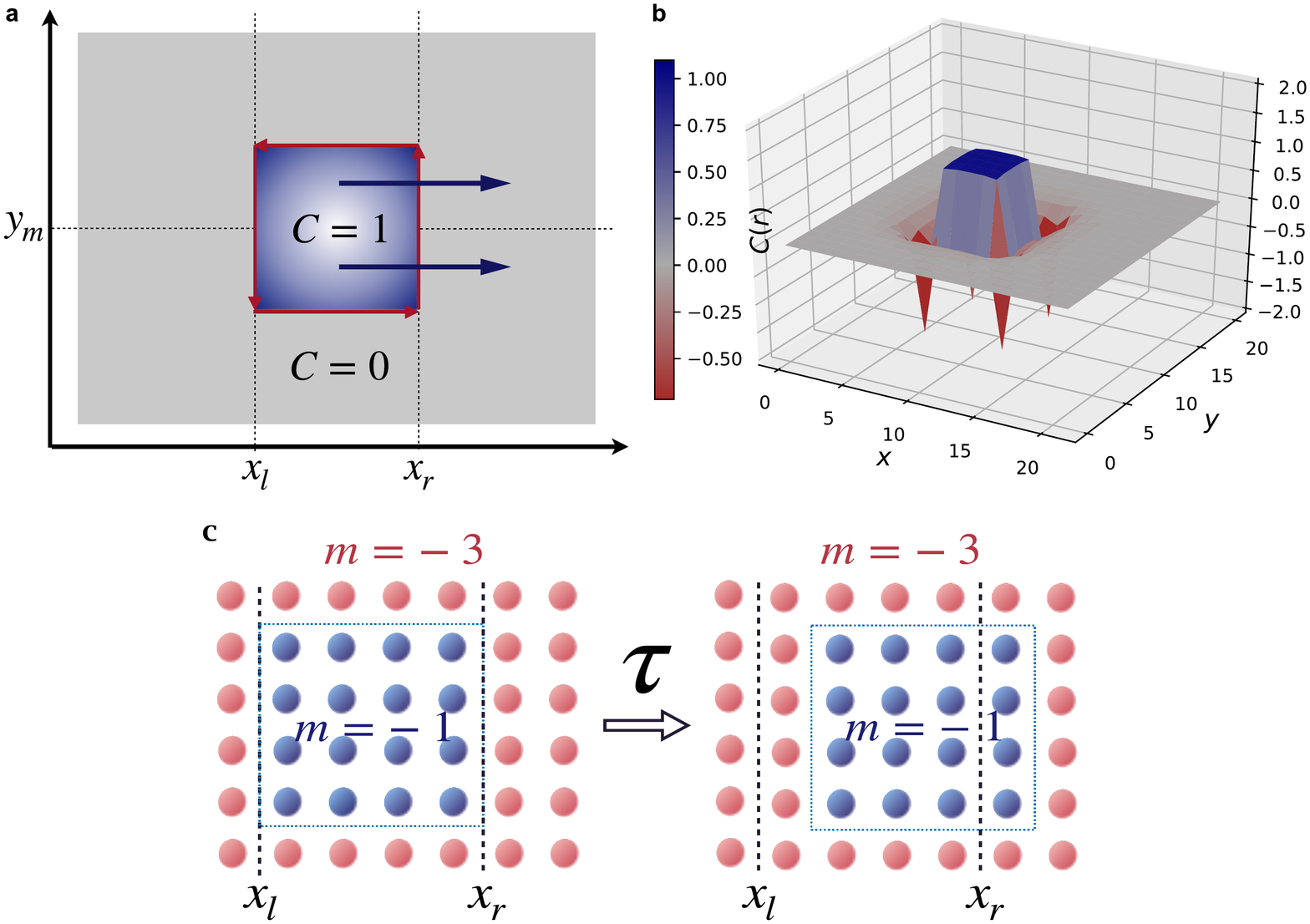"}
	\caption{\textbf{Scheme of the setting} \textbf{a} The domain with ''Chern number'' $C=1$ (blue in figure) is drifting inside a topologically trivial sample (grey $C=0$). The dark blue arrows indicate the drift direction of the topologically non-trivial domain. \textbf{b} The initial distribution of the local Chern  \cite{bianco2011mapping}. \textbf{c} The evolution of the on-site magnetic field $m$, controlling the phase in the model~\ref{eq:Ch_eq}. }
	\label{fig:setting}
\end{figure}

The space and time dependency of $m$ corresponds to the slow drift of a topological domain inside a system in a trivial phase (see Fig.~\ref{fig:setting}). Different phases are created by selecting the parameter $m$ as dictated by the Eq. (\ref{m_det}). In our case $m_1=-3$ and $m_2=-1$ were chosen for trivial and topological phase correspondingly. Initially the system is prepared in the ground state of $H(0)$. For the domain to move, we change the parameter $m$ at the boundaries of two different phases on the lattice from $m_1$ to $m_2$. The time protocol for such process can be expressed as follows:
\begin{align}  \label{eq:m_par}
m(x, t)= \left  \{ \begin{array}{lllll}
m_1, & x<x_l+[t / \tau] \\
m_2(1-t/\tau)+m_1\cdot t/\tau & x=x_l+[t/\tau] \\
m_2, & x_l+[t/\tau]< x< x_r+[t/\tau] \\
m_1(1-t/\tau)+m_2\cdot t/\tau & x=x_r+[t/\tau] \\
m_1, & x>x_r+[t/\tau] 
\end{array} \right.
\end{align} 
where $x_l$, $x_r$ denote two sites on the left and right border at the beginning of time protocol (See Fig.~\ref{fig:setting}.) and $\tau$ determine the period of protocol when the domain shifts on one site. \\

\textit{Unitary evolution.}
The evolution of the system is governed by the time-ordered exponent:
\begin{equation} \label{eq:unit_evol}
| \psi (t) \rangle = T\left\{\exp \left(-\frac{i}{\hbar} \int_{t_{0}}^{t} H\left(t^{\prime}\right) d t^{\prime}\right)\right\} | \psi (t_0) \rangle
\end{equation}
Numerically we approximated the state $| \psi (t) \rangle$ using the Trotter-Suzuki decomposition \cite{trotter1959product,suzuki1976generalized}. 

\textit{Electric current.} For detecting edge modes we used the following operator for electrical currents
\begin{equation}
\hat{J}(\bm{r}_2, \bm{r}_2)= i \sum_{\sigma,\sigma'}\left( c_{\bm{r}_2, \sigma}^{\dagger} t(\bm{r}_1,\bm{r}_2) c_{\bm{r}_1, \sigma'}-\text { h.c. }\right) .
\end{equation}




\section*{Acknowledgements}

We thank Peru d’Ornellas for many insightful discussions and in particular for sharing the details of the ongoing work \cite{peru}. Useful comments from Oleg Dubinkin are gratefully acknowledged. This work was carried out in the framework of the Russian Quantum Technologies Roadmap. A.A.M. was also also supported by the “Basis” foundation under Grant No. 18-3-01

\section*{Author contributions statement}

D.B.G. and A.R.Y. performed the numerical simulations. A.A.M. initiated and directed the project. Eq.~\ref{eq:nq_streda} is due to A.N.R.  All the authors contributed to the results analysis and writing the manuscript.

\end{document}